\documentclass{article}

\PassOptionsToPackage{numbers,sort&compress}{natbib}

\usepackage[final]{nips_2016}

\usepackage[utf8]{inputenc} 
\usepackage[T1]{fontenc}    
\usepackage{hyperref}       
\usepackage{url}            
\usepackage{booktabs}       
\usepackage{amsfonts}       
\usepackage{nicefrac}       
\usepackage{microtype}      
\usepackage{bm}
\usepackage{amsmath}
\usepackage{graphicx}
\usepackage[ruled]{algorithm2e}
\usepackage{amssymb}
\usepackage[margin=.5cm]{caption}
\usepackage{subcaption}

\DeclareMathOperator*{\argmax}{arg \, max}
\DeclareMathOperator*{\argmin}{arg \, min}

\title{Maximizing Influence in an Ising Network:\\
A Mean-Field Optimal Solution}

\author{
  Christopher W. Lynn \\
  Department of Physics and Astronomy \\
  University of Pennsylvania \\
  \texttt{chlynn@sas.upenn.edu}
  \And
  Daniel D. Lee \\
  Department of Electrical and Systems Engineering \\
  University of Pennsylvania \\
  \texttt{ddlee@seas.upenn.edu}
}

\begin{document}

\maketitle

\begin{abstract}

Influence maximization in social networks has typically been studied in the context of contagion models and irreversible processes. In this paper, we consider an alternate model that treats individual opinions as spins in an Ising system at dynamic equilibrium. We formalize the \textit{Ising influence maximization} problem, which has a natural physical interpretation as maximizing the magnetization given a budget of external magnetic field. Under the mean-field (MF) approximation, we present a gradient ascent algorithm that uses the susceptibility to efficiently calculate local maxima of the magnetization, and we develop a number of sufficient conditions for when the MF magnetization is concave and our algorithm converges to a global optimum. We apply our algorithm on random and real-world networks, demonstrating, remarkably, that the MF optimal external fields (i.e., the external fields which maximize the MF magnetization) shift from focusing on high-degree individuals at high temperatures to focusing on low-degree individuals at low temperatures. We also establish a number of novel results about the structure of steady-states in the ferromagnetic MF Ising model on general graph topologies, which are of independent interest.

\end{abstract}

\section{Introduction}

With the proliferation of online social networks, the problem of optimally influencing the opinions of individuals in a population has garnered tremendous attention \cite{Domingos-01, Richardson-01, Kempe-01}. The prevailing paradigm treats marketing as a viral process, whereby the advertiser is given a budget of seed infections and chooses the subset of individuals to infect such that the spread of the ensuing contagion is maximized. The development of algorithmic methods for influence maximization under the viral paradigm has been the subject of vigorous study, resulting in a number of efficient techniques for identifying meaningful marketing strategies in real-world settings \cite{Mossel-01,Goyal-01, Gomez-Rodriguez-01}. While the viral paradigm accurately describes out-of-equilibrium phenomena, such as the introduction of new ideas or products to a system, these models fail to capture reverberant opinion dynamics wherein repeated interactions between individuals in the network give rise to complex macroscopic opinion patterns, as, for example, is the case in the formation of political opinions \cite{Galam-03,Isenberg-01,Mas-01,Moussaid-01}. In this context, rather than maximizing the spread of a viral advertisement, the marketer is interested in optimally shifting the equilibrium opinions of individuals in the network.

To describe complex macroscopic opinion patterns resulting from repeated microscopic interactions, we naturally employ the language of statistical mechanics, treating individual opinions as spins in an Ising system at dynamic equilibrium and modeling marketing as the addition of an external magnetic field. The resulting problem, which we call \textit{Ising influence maximization (IIM)}, has a natural physical interpretation as maximizing the magnetization of an Ising system given a budget of external field. While a number of models have been proposed for describing reverberant opinion dynamics \cite{De-01}, our use of the Ising model follows a vibrant interdisciplinary literature \cite{Montanari-01,Castellano-01}, and is closely related to models in game theory \cite{Blume-01,McKelvey-01} and sociophysics \cite{Galam-02,Sznajd-01}. Furthermore, complex Ising models have found widespread use in machine learning, and our model is formally equivalent to a pair-wise Markov random field or a Boltzmann machine \cite{Kindermann-01, Tanaka-01, Nishimori-01}.

Our main contributions are as follows:
\begin{enumerate}
\item We formalize the influence maximization problem in the context of the Ising model, which we call the \textit{Ising influence maximization (IIM)} problem. We also propose the \textit{mean-field Ising influence maximization (MF-IIM)} problem as an approximation to IIM (Section 2).
\item We find sufficient conditions under which the MF-IIM objective is smooth and concave, and we present a gradient ascent algorithm that guarantees an $\epsilon$-approximation to MF-IIM (Section 4).
\item We present numerical simulations that probe the structure and performance of MF optimal marketing strategies. We find that at high temperatures, it is optimal to focus influence on high-degree individuals, while at low temperatures, it is optimal to spread influence among low-degree individuals (Sections 5 and 6).
\item Throughout the paper we present a number of novel results concerning the structure of steady-states in the ferromagnetic MF Ising model on general (weighted, directed) strongly-connected graphs, which are of independent interest. We name two highlights:
\begin{itemize}
\item The well-known pitchfork bifurcation structure for the ferromagnetic MF Ising model on a lattice extends exactly to general strongly-connected graphs, and the critical temperature is equal to the spectral radius of the adjacency matrix (Theorem 3).
\item There can exist at most one stable steady-state with non-negative (non-positive) components, and it is smooth and concave (convex) in the external field (Theorem 4). 
\end{itemize}
\end{enumerate}

\section{The Ising influence maximization problem}

We consider a weighted, directed social network consisting of a set of individuals $N=\{1,\hdots,n\}$, each of which is assigned an opinion $\sigma_i\in\{\pm 1\}$ that captures its current state. By analogy with the Ising model, we refer to $\bm{\sigma}=(\sigma_i)$ as a spin configuration of the system. Individuals in the network interact via a non-negative weighted coupling matrix $J\in \mathbb{R}^{n\times n}_{\ge 0}$, where $J_{ij}\ge 0$ represents the amount of influence that individual $j$ holds over the opinion of individual $i$, and the non-negativity of $J$ represents the assumption that opinions of neighboring individuals tend to align, known in physics as a ferromagnetic interaction. Each individual also interacts with forces external to the network via an external field $\bm{h}\in\mathbb{R}^n$. For example, if the spins represent the political opinions of individuals in a social network, then $J_{ij}$ represents the influence that $j$ holds over $i$'s opinion and $h_i$ represents the political bias of node $i$ due to external forces such as campaign advertisements and news articles.

The opinions of individuals in the network evolve according to asynchronous Glauber dynamics. At each time $t$, an individual $i$ is selected uniformly at random and her opinion is updated in response to the external field $\bm{h}$ and the opinions of others in the network $\bm{\sigma}(t)$ by sampling from
\begin{equation}
\label{Glauber}
P\left(\sigma_i(t+1) = 1|\bm{\sigma}(t)\right) = \frac{e^{\beta\left(\sum_j J_{ij}\sigma_j(t) + h_i\right)}}{\sum_{\sigma'_i=\pm 1} e^{\beta \sigma'_i\left(\sum_j J_{ij}\sigma_j(t) + h_i\right)}},
\end{equation}
where $\beta$ is the inverse temperature, which we refer to as the \textit{interaction strength}, and unless otherwise specified, sums are assumed over $N$. Together, the quadruple $(N,J, \bm{h},\beta)$ defines our system. We refer to the total expected opinion, $M=\sum_i\left< \sigma_i\right>$, as the \textit{magnetization}, where $\left<\cdot\right>$ denotes an average over the dynamics in Eq. (\ref{Glauber}), and we often consider the magnetization as a function of the external field, denoted $M(\bm{h})$. Another important concept is the \textit{susceptibility} matrix, $\chi_{ij} = \frac{\partial \left<\sigma_i\right>}{\partial h_j}$, which quantifies the response of individual $i$ to a change in the external field on node $j$.

We study the problem of maximizing the magnetization of an Ising system with respect to the external field. We assume that an external field $\bm{h}$ can be added to the system, subject to the constraints $\bm{h}\ge 0$ and $\sum_i h_i\le H$, where $H>0$ is the \textit{external field budget}, and we denote the set of feasible external fields by $\mathcal{F}_H= \left\{\bm{h}\in\mathbb{R}^n : \bm{h}\ge 0, \sum_i h_i = H \right\}$. In general, we also assume that the system experiences an initial external field $\bm{b}\in \mathbb{R}^n$, which cannot be controlled.

\textbf{Definition 1.} (\textit{Ising influence maximization (IIM)}) Given a system $(N, J, \bm{b},\beta)$ and a budget $H$, find a feasible external field $\bm{h}\in \mathcal{F}_H$ that maximizes the magnetization; that is, find an optimal external field $\bm{h}^*$ such that
\begin{equation}
\bm{h}^* = \argmax_{\bm{h}\in\mathcal{F}_H} M(\bm{b}+\bm{h}).
\end{equation}

\paragraph{Notation.} Unless otherwise specified, bold symbols represent column vectors with the appropriate number of components, while non-bold symbols with subscripts represent individual components. We often abuse notation and write relations such as $\bm{m}\ge 0$ to mean $m_i\ge 0$ for all components $i$.

\subsection{The mean-field approximation}

In general, calculating expectations over the dynamics in Eq. (\ref{Glauber}) requires Monte-Carlo simulations or other numerical approximation techniques. To make analytic progress, we employ the variational mean-field approximation, which has roots in statistical physics and has long been used to tackle inference problems in Boltzmann machines and Markov random fields \cite{Yedidia-01,Jordan-01,Opper-01,Saul-01}. The mean-field approximation replaces the intractable task of calculating exact averages over Eq. (\ref{Glauber}) with the problem of solving the following set of self-consistency equations:
\begin{equation}
\label{MF}
m_i = \tanh\left[\beta\left(\sum_j J_{ij}m_j + h_i\right)\right],
\end{equation}

for all $i\in N$, where $m_i$ approximates $\left<\sigma_i\right>$. We refer to the right-hand side of Eq. (\ref{MF}) as the \textit{mean-field map}, $\bm{f}(\bm{m})=\tanh\left[\beta(J\bm{m}+\bm{h})\right]$, where $\tanh(\cdot)$ is applied component-wise. In this way, a fixed point of the mean-field map is a solution to Eq. (\ref{MF}), which we call a \textit{steady-state}. 

In general, there may be many solutions to Eq. (\ref{MF}), and we denote by $\mathcal{M}_{\bm{h}}$ the set of steady-states for a system $(N,J,\bm{h},\beta)$. We say that a steady-state $\bm{m}$ is \textit{stable} if $\rho(\bm{f}'(\bm{m}))<1$, where $\rho(\cdot)$ denotes the spectral radius and
\begin{equation}
\bm{f}'(\bm{m})_{ij} = \left.\frac{\partial f_i}{\partial m_j}\right\vert_{\bm{m}} = \beta\left(1-m_i^2\right)J_{ij} \quad \Rightarrow \quad \bm{f}'(\bm{m}) = \beta D(\bm{m}) J,
\end{equation}
where $D(\bm{m})_{ij} = (1-m_i^2)\delta_{ij}$. Furthermore, under the mean-field approximation, given a stable steady-state $\bm{m}$, the susceptibility has a particularly nice form:
\begin{equation}
\chi_{ij}^{MF} = \beta\left(1-m_i^2\right)\left(\sum_k J_{ik}\chi_{kj} + \delta_{ij}\right) \quad \Rightarrow \quad \chi^{MF} = \beta\left(I-\beta D(\bm{m})J\right)^{-1}D(\bm{m}),
\end{equation}
where $I$ is the $n\times n$ identity matrix.

For the purpose of uniquely defining our objective, we optimistically choose to maximize the maximum magnetization among the set of steady-states, defined by
\begin{equation}
M^{MF}(\bm{h}) = \max_{\bm{m}\in\mathcal{M}_{\bm{h}}} \sum_i m_i(\bm{h}).
\end{equation}
We note that the pessimistic framework of maximizing the minimum magnetization yields an equally valid objective. We also note that simply choosing a steady-state to optimize does not yield a well-defined objective since, as $\bm{h}$ increases, steady-states can pop in and out of existence.

\textbf{Definition 2.} (\textit{Mean-field Ising influence maximization (MF-IIM)}) Given a system $(N, J, \bm{b},\beta)$ and a budget $H$, find an optimal external field $\bm{h}^*$ such that
\begin{equation}
\bm{h}^* = \argmax_{\bm{h}\in\mathcal{F}_H} M^{MF}(\bm{b}+\bm{h}).
\end{equation}

\section{The structure of steady-states in the MF Ising model}

Before proceeding further, we must prove an important result concerning the existence and structure of solutions to Eq. (\ref{MF}), for if there exists a system that does not admit a steady-state, then our objective is ill-defined. Furthermore, if there exists a unique steady-state $\bm{m}$, then $M^{MF}=\sum_i m_i$, and there is no ambiguity in our choice of objective.

Theorem 3 establishes that every system admits a steady-state and that the well-known pitchfork bifurcation structure for steady-states of the ferromagnetic MF Ising model on a lattice extends exactly to general (weighted, directed) strongly-connected graphs. In particular, for any strongly-connected graph described by $J$, there is a \textit{critical interaction strength}  $\beta_c$ below which there exists a unique and stable steady-state. For $\bm{h}=0$, as $\beta$ crosses $\beta_c$ from below, two new stable steady-states appear, one with all-positive components and one with all-negative components. Interestingly, the critical interaction strength is equal to the inverse of the spectral radius of $J$, denoted $\beta_c=1/\rho(J)$.

\textbf{Theorem 3.}
\textit{Any system $(N,J,\bm{h},\beta)$ exhibits a steady-state. Furthermore, if its network is strongly-connected, then, for $\beta< \beta_c$, there exists a unique and stable steady-state. For $\bm{h}=0$, as $\beta$ crosses $\beta_c$ from below, the unique steady-state gives rise to two stable steady-states, one with all-positive components and one with all-negative components.}

\textit{Proof sketch.} 
The existence of a steady-state follows directly by applying Brouwer's fixed-point theorem to $\bm{f}$. For $\beta<\beta_c$, it can be shown that $\bm{f}$ is a contraction mapping, and hence admits a unique and stable steady-state by Banach's fixed point theorem. For $\bm{h}=0$ and $\beta<\beta_c$, $\bm{m}=0$ is the unique steady-state and $\bm{f}'(\bm{m})=\beta J$. Because $J$ is strongly-connected, the Perron-Frobenius theorem guarantees a simple eigenvalue equal to $\rho(J)$ and a corresponding all-positive eigenvector. Thus, when $\beta$ crosses $1/\rho(J)$ from below, the Perron-Frobenius eigenvalue of $\bm{f}'(\bm{m})$ crosses 1 from below, giving rise to a supercritical pitchfork bifurcation with two new stable steady-states corresponding to the Perron-Frobenius eigenvector.

\textit{Remark.}
Some of our results assume $J$ is strongly-connected in order to use the Perron-Frobenius theorem. We note that this assumption is not restrictive, since any graph can be efficiently decomposed into strongly-connected components on which our results apply independently.

Theorem 3 shows that the objective $M^{MF}(\bm{b}+\bm{h})$ is well-defined. Furthermore, for $\beta<\beta_c$, Theorem 3 guarantees a unique and stable steady-state $\bm{m}$ for all $\bm{b}+\bm{h}$. In this case, MF-IIM reduces to maximizing $M^{MF}=\sum_i m_i$, and because $\bm{m}$ is stable, $M^{MF}(\bm{b}+\bm{h})$ is smooth for all $\bm{h}$ by the implicit function theorem. Thus, for $\beta<\beta_c$, we can use standard gradient ascent techniques to efficiently calculate locally-optimal solutions to MF-IIM. In general, $M^{MF}$ is not necessarily smooth in $\bm{h}$ since the topological structure of steady-states may change as $\bm{h}$ varies. However, in the next section we show that if there exists a stable and entry-wise non-negative steady-state, and if $J$ is strongly-connected, then $M^{MF}(\bm{b}+\bm{h})$ is both smooth and concave in $\bm{h}$, regardless of the interaction strength.

\section{Sufficient conditions for when MF-IIM is concave}

We consider conditions for which MF-IIM is smooth and concave, and hence exactly solvable by efficient techniques. The case under consideration is when $J$ is strongly-connected and there exists a stable non-negative steady-state.

\textbf{Theorem 4.}
\textit{Let $(N,J,\bm{b},\beta)$ describe a system with a strongly-connected graph for which there exists a stable non-negative steady-state $\bm{m}(\bm{b})$. Then, for any $H$, $M^{MF}(\bm{b}+\bm{h})=\sum_i m_i(\bm{b}+\bm{h})$, $M^{MF}(\bm{b}+\bm{h})$ is smooth in $\bm{h}$, and $M^{MF}(\bm{b}+\bm{h})$ is concave in $\bm{h}$ for all $\bm{h}\in\mathcal{F}_H$.}

\textit{Proof sketch.} 
Our argument follows in three steps. We first show that $\bm{m}(\bm{b})$ is the unique stable non-negative steady-state and that it attains the maximum total opinion among steady-states. This guarantees that $M^{MF}(\bm{b})=\sum_i m_i(\bm{b})$. Furthermore, $\bm{m}(\bm{b})$ gives rise to a unique and smooth branch of stable non-negative steady-states for additional $\bm{h}$, and hence $M^{MF}(\bm{b}+\bm{h})=\sum_i m_i(\bm{b}+\bm{h})$ for all $\bm{h}>0$.  Finally, one can directly show that $M^{MF}(\bm{b}+\bm{h})$ is concave in $\bm{h}$.

\textit{Remark.}
By arguments similar to those in Theorem 4, it can be shown that any stable non-positive steady-state is unique, attains the minimum total opinion among steady-states, and is smooth and convex for decreasing $\bm{h}$.

The above result paints a significantly simplified picture of the MF-IIM problem when $J$ is strongly-connected and there exists a stable non-negative steady-state $\bm{m}(\bm{b})$. Given a budget $H$, for any feasible marketing strategy $\bm{h}\in \mathcal{F}_H$, $\bm{m}(\bm{b}+\bm{h})$ is the unique stable non-negative steady-state, attains the maximum total opinion among steady-states, and is smooth in $\bm{h}$. Thus, the objective $M^{MF}(\bm{b}+\bm{h})=\sum_i m_i(\bm{b}+\bm{h})$ is smooth, allowing us to write down a gradient ascent algorithm that approximates a local maximum. Furthermore, since $M^{MF}(\bm{b}+\bm{h})$ is concave in $\bm{h}$, any local maximum of $M^{MF}$ on $\mathcal{F}_H$ is a global maximum, and we can apply efficient gradient ascent techniques to solve MF-IIM. 

Our algorithm, summarized in Algorithm 1, is initialized at a feasible external field. At each iteration, we calculate the susceptibility of the system, namely $\frac{\partial M^{MF}}{\partial h_j} = \sum_i \chi^{MF}_{ij}$, and project this gradient onto $\mathcal{F}_H$ (the projection operator $P_{\mathcal{F}_H}$ is well-defined since $\mathcal{F}_H$ is convex). Stepping along the direction of the projected gradient with step size $\alpha\in (0,\frac{1}{L})$, where $L$ is a Lipschitz constant of $M^{MF}$, Algorithm 1 converges to an $\epsilon$-approximation to MF-IIM in $O(1/\epsilon)$ iterations \cite{Teboulle-01}.
\begin{algorithm}[t]
\KwIn{System $(N,J,\bm{b},\beta)$ for which there exists a stable non-negative steady-state, budget $H$, accuracy parameter $\epsilon>0$}
\KwOut{External field $\bm{h}$ that approximates a MF optimal external field $\bm{h}^*$}
$t=0$; $\bm{h}(0)\in\mathcal{F}_H$; $\alpha \in (0,\frac{1}{L})$ \;
\Repeat{$M^{MF}(\bm{b}+\bm{h}^*) - M^{MF}(\bm{b}+\bm{h}(t)) \le \epsilon$}{
$\frac{\partial M^{MF}(\bm{b}+\bm{h}(t))}{\partial h_j} = \sum_i \chi_{ij}^{MF}(\bm{b}+\bm{h}(t))$\;
$\bm{h}(t+1) = P_{\mathcal{F}_H}\left[\bm{h}(t) + \alpha \triangledown_{\bm{h}}M^{MF}(\bm{b}+\bm{h}(t))\right]$\;
$t$++\;
}
$\bm{h}= \bm{h}(t)$\;
\caption{An $\epsilon$-approximation to MF-IIM}
\label{alg:one}
\end{algorithm}

\subsection{Sufficient conditions for the existence of a stable non-negative steady-state}

In the previous section we found that MF-IIM is efficiently solvable if there exists a stable non-negative steady-state. While this assumption may seem restrictive, we show, to the contrary, that the appearance of a stable non-negative steady-state is a fairly general phenomenon. We first show, for $J$ strongly-connected, that the existence of a stable non-negative steady-state is robust to increases in $\bm{h}$ and that the existence of a stable positive steady-state is robust to increases in $\beta$.

\textbf{Theorem 5.}
\textit{Let $(N,J,\bm{h},\beta)$ describe a system with a strongly-connected graph for which there exists a stable non-negative steady-state $\bm{m}$. If $\bm{m}\ge 0$, then as $\bm{h}$ increases, $\bm{m}$ gives rise to a unique and smooth branch of stable non-negative steady-states. If $\bm{m}> 0$, then as $\beta$ increases, $\bm{m}$ gives rise to a unique and smooth branch of stable positive steady-states.}

\textit{Proof sketch.}
By the implicit function theorem, any stable steady-state can be locally defined as a function of both $\bm{h}$ and $\beta$. Using the susceptibility, one can directly show that any stable non-negative steady-state remains stable and non-negative as $\bm{h}$ increases and that any stable positive steady-state remains stable and positive as $\beta$ increases.

The intuition behind Theorem 5 is that increasing the external field will never destroy a steady-state in which all of the opinions are already non-positive. Furthermore, as the interaction strength increases, each individual reacts more strongly to the positive influence of her neighbors, creating a positive feedback loop that results in an even more positive magnetization. We conclude by showing for $J$ strongly-connected that if $\bm{h}\ge 0$, then there exists a stable non-negative steady-state.

\textbf{Theorem 6.}
\textit{Let $(N,J,\bm{h},\beta)$ describe any system with a strongly-connected network. If $\bm{h}\ge 0$, then there exists a stable non-negative steady-state.}

\textit{Proof sketch.} For $\bm{h}>0$ and $\beta<\beta_c$, it can be shown that the unique steady-state is positive, and hence Theorem 5 guarantees the result for all $\beta'>\beta$. For $\bm{h}=0$, Theorem 3 provides the result.

All together, the results of this section provide a number of sufficient conditions under which MF-IIM is exactly and efficiently solvable by Algorithm 1.

\section{A shift in the structure of solutions to MF-IIM}

The structure of solutions to MF-IIM is of fundamental theoretical and practical interest. We demonstrate, remarkably, that solutions to MF-IIM shift from focusing on nodes of high degree at low interaction strengths to focusing on nodes of low degree at high interaction strengths.

Consider an Ising system described by $(N,J,\bm{h},\beta)$ in the limit $\beta\ll \beta_c$. To first-order in $\beta$, the self-consistency equations (\ref{MF}) take the form:
\begin{equation}
\bm{m} = \beta\left(J\bm{m}+\bm{h}\right) \quad \Rightarrow\quad \bm{m} = \beta(I-\beta J)^{-1}\bm{h}.
\end{equation}
Since $\beta<\beta_c$, we have $\rho(\beta J) < 1$, allowing us to expand $(I-\beta J)^{-1}$ in a geometric series:
\begin{equation}
\bm{m} = \beta\bm{h} + \beta^2J\bm{h} + O(\beta^3)\quad \Rightarrow \quad M^{MF}(\bm{h}) = \beta\sum_i h_i + \beta^2\sum_i d^{out}_i h_i + O(\beta^3),
\end{equation}
where $d^{out}_i = \sum_j J_{ji}$ is the out-degree of node $i$. Thus, for low interaction strengths, the MF magnetization is maximized by focusing the external field on the nodes of highest out-degree in the network, independent of $\bm{b}$ and $H$.

To study the structure of solutions to MF-IIM at high interaction strengths, we make the simplifying assumptions that $J$ is strongly-connected and $\bm{b}\ge 0$ so that Theorem 6 guarantees a stable non-negative steady state $\bm{m}$. For large $\beta$ and an additional external field $\bm{h}\in\mathcal{F}_H$, $\bm{m}$ takes the form
\begin{equation}
m_i \approx \tanh\left[\beta\left(\sum_j J_{ij} + b_i + h_i\right)\right] \approx 1 - 2e^{-2\beta(d^{in}_i + b_i + h_i)},
\end{equation}
where $d^{in}_i=\sum_j J_{ij}$ is the in-degree of node $i$. Thus, in the high-$\beta$ limit, we have:
\begin{equation}
M^{MF}(\bm{b}+\bm{h}) \approx  \sum_i \left( 1 - 2e^{-2\beta(d^{in}_i + b_i + h_i)}\right) \approx n - 2e^{-2\beta(d^{in}_{i^*} + h_{i^*}^{(0)} + h_{i^*})},
\end{equation}
where $i^*=\argmin_i(d^{in}_i + b_i + h_i)$. Thus, for high interaction strengths, the solutions to MF-IIM for an external field budget $H$ are given by:
\begin{equation}
\label{lowT}
\bm{h}^* = \argmax_{\bm{h}\in\mathcal{F}_H} \left(n - 2e^{-2\beta(d^{in}_{i^*} + h_{i^*}^{(0)} + h_{i^*})}\right) \equiv \argmax_{\bm{h}\in\mathcal{F}_H} \min_i\left( d^{in}_i + b_i + h_i \right).
\end{equation}
Eq. (\ref{lowT}) reveals that the high-$\beta$ solutions to MF-IIM focus on the nodes for which $ d^{in}_i + b_i + h_i$ is smallest. Thus, if $\bm{b}$ is uniform, the MF magnetization is maximized by focusing the external field on the nodes of smallest in-degree in the network.

We emphasize the strength and novelty of the above results. In the context of reverberant opinion dynamics, the optimal control strategy has a highly non-trivial dependence on the strength of interactions in the system, a feature not captured by viral models.  Thus, when controlling a social system, accurately determining the strength of interactions is of critical importance.

\section{Numerical simulations}

We present numerical experiments to probe the structure and performance of MF optimal external fields. We verify that the solutions to MF-IIM undergo a shift from focusing on high-degree nodes at low interaction strengths to focusing on low-degree nodes at high interaction strengths. We also find that for sufficiently high and low interaction strengths, the MF optimal external field achieves the maximum exact magnetization, while admitting performance losses near $\beta_c$. However, even at $\beta_c$, we demonstrate that solutions to MF-IIM significantly outperform common node-selection heuristics based on node degree and centrality.

We first consider an undirected hub-and-spoke network, shown in Figure 1, where $J_{ij}\in \{0,1\}$ and we set $\bm{b}=0$ for simplicity. Since $\bm{b}\ge 0$, Algorithm 1 is guaranteed to achieve a globally optimal MF magnetization. Furthermore, because the network is small, we can calculate exact solutions to IIM by brute force search. The left plot in Figure 1 compares the average degree of the MF and exact optimal external fields over a range of temperatures for an external field budget $H=1$, verifying that the solution to MF-IIM shifts from focusing on high-degree nodes at low interaction strengths to low-degree nodes at high interaction strengths. Furthermore, we find that the shift in the MF optimal external field occurs near the critical interaction strength $\beta_c = .5$. The performance of the MF optimal strategy (measured as the ratio of the magnetization achieved by the MF solution to that achieved by the exact solution) is shown in the right plot in Figure 1. For low and high interaction strengths, the MF optimal external field achieves the maximum magnetization, while near $\beta_c$, it incurs significant performance losses, a phenomenon well-studied in the literature \cite{Yedidia-01}.
\begin{figure}[t]
\centering
 \begin{subfigure}{0.4\textwidth}
        \includegraphics[width=\linewidth]{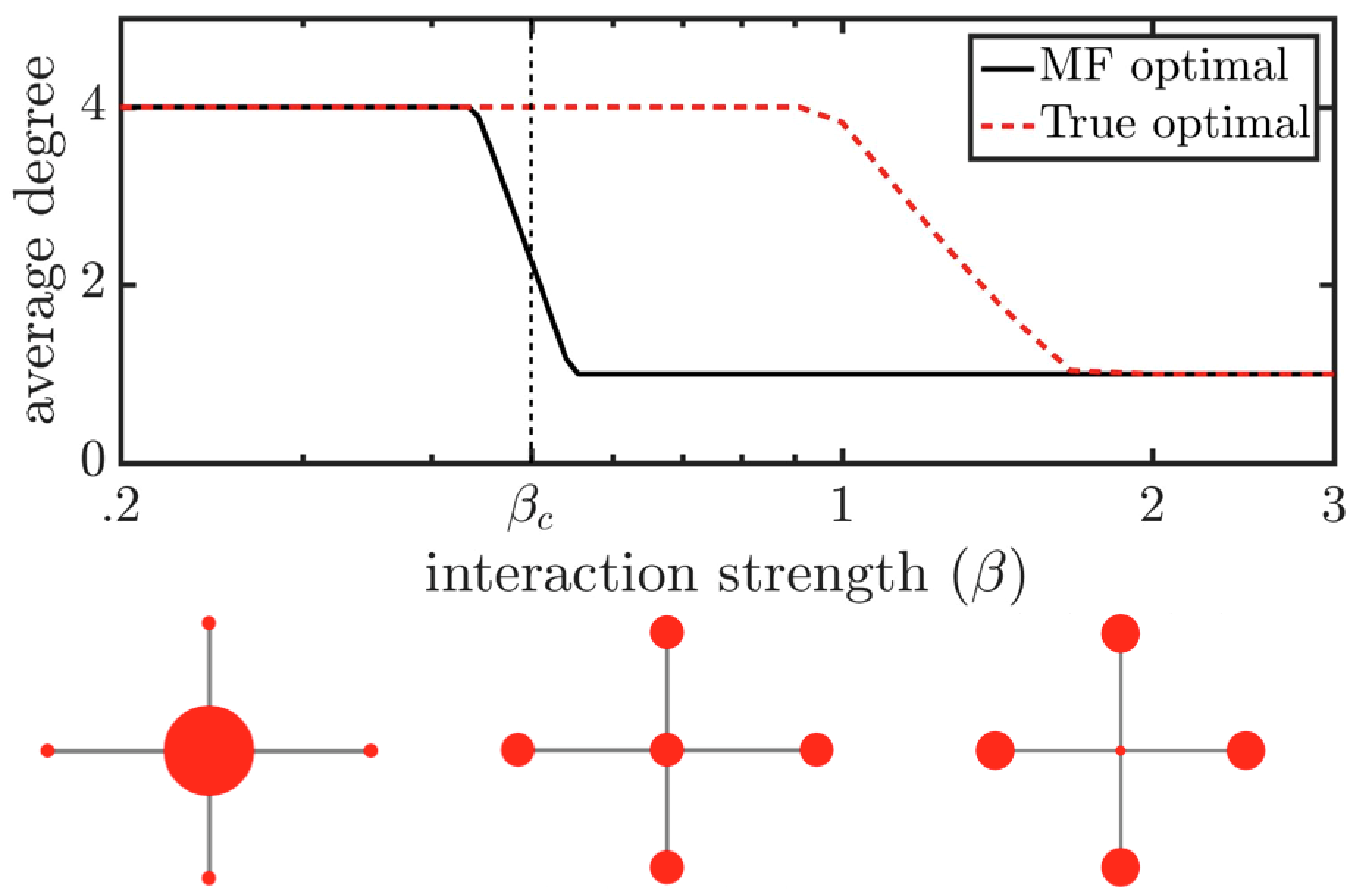}
\end{subfigure}%
\begin{subfigure}{0.05\textwidth}
\includegraphics[width=\linewidth]{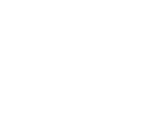}
\end{subfigure}%
\begin{subfigure}{0.4\textwidth}
        \includegraphics[width=\linewidth]{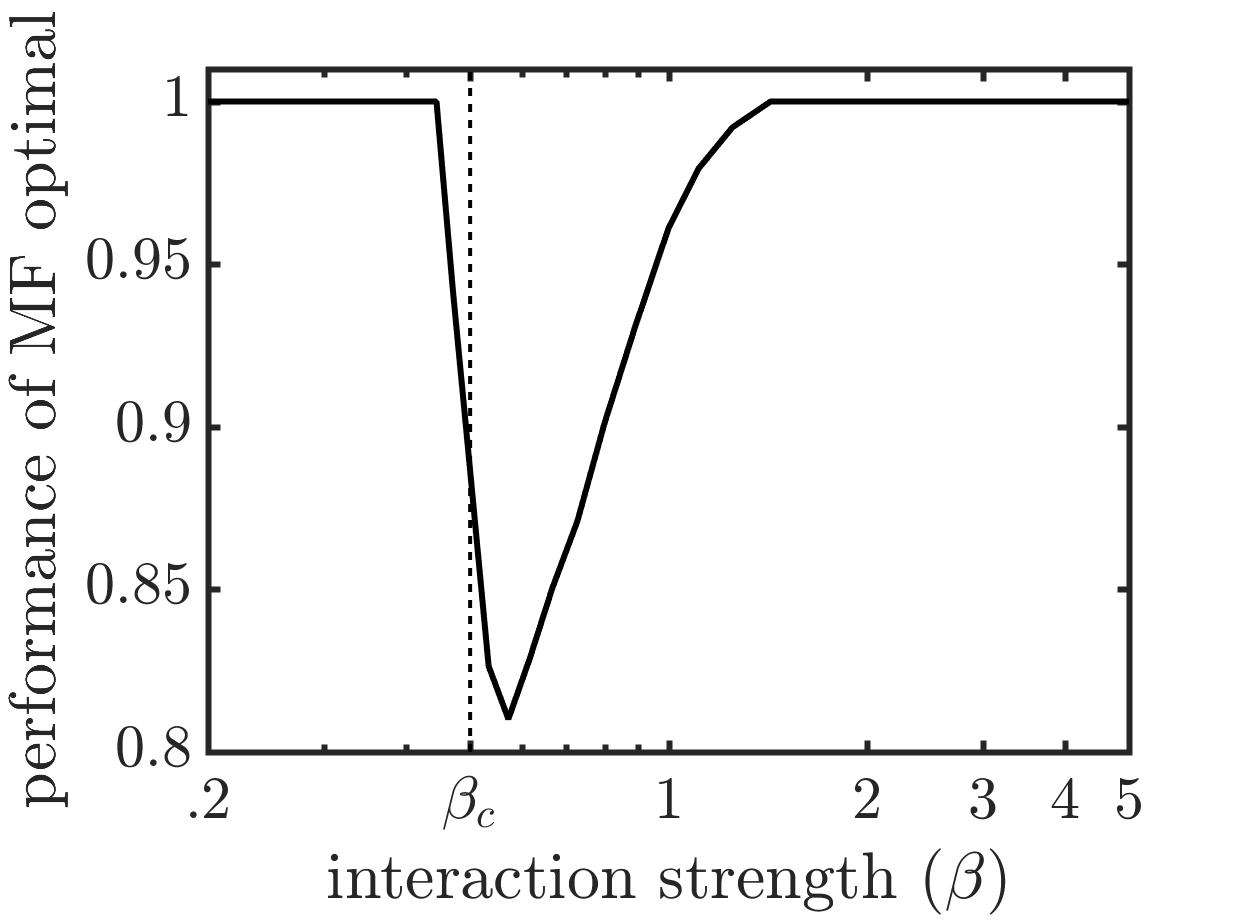}
\end{subfigure}
\caption{Left: A comparison of the structure of the MF and exact optimal external fields, denoted $\bm{h}^*_{MF}$ and $\bm{h}^*$, in a hub-and-spoke network. Right: The relative performance of $\bm{h}^*_{MF}$ compared to $\bm{h}^*$; i.e., $M(\bm{h}^*_{MF})/M(\bm{h}^*_{MF})$, where $M$ denotes the exact magnetization.}
\end{figure}

We now consider a stochastic block network consisting of 100 nodes split into two blocks of 50 nodes each, shown in Figure 2. An undirected edge of weight 1 is placed between each pair of nodes in Block 1 with probability .2, between each pair in Block 2 with probability .05, and between nodes in different blocks with probability .05, resulting in a highly-connected community (Block 1) surrounded by a sparsely-connected community (Block 2). For $\bm{b}=0$ and $H=20$, the center plot in Figure 2 demonstrates that the solution to MF-IIM shifts from focusing on Block 1 at low $\beta$ to focusing on Block 2 at high $\beta$ and that the shift occurs near $\beta_c$.
\begin{figure}[t]
\centering
 \begin{subfigure}{0.2\textwidth}
        \includegraphics[width=\linewidth]{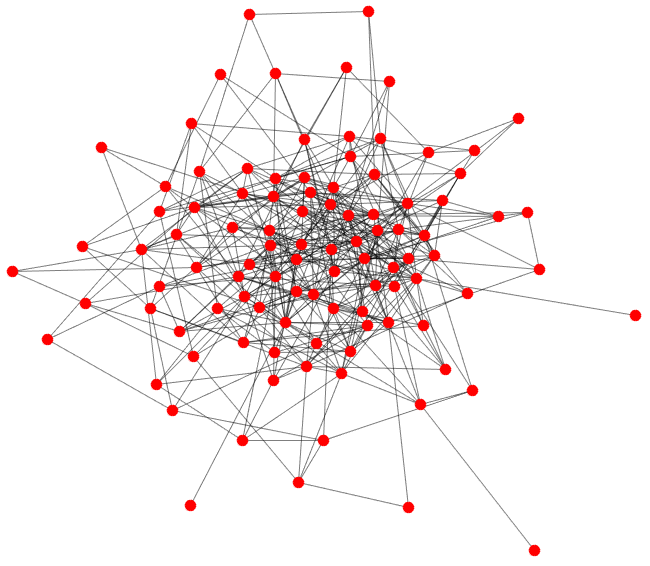}
\end{subfigure}%
\begin{subfigure}{0.4\textwidth}
        \includegraphics[width=\linewidth]{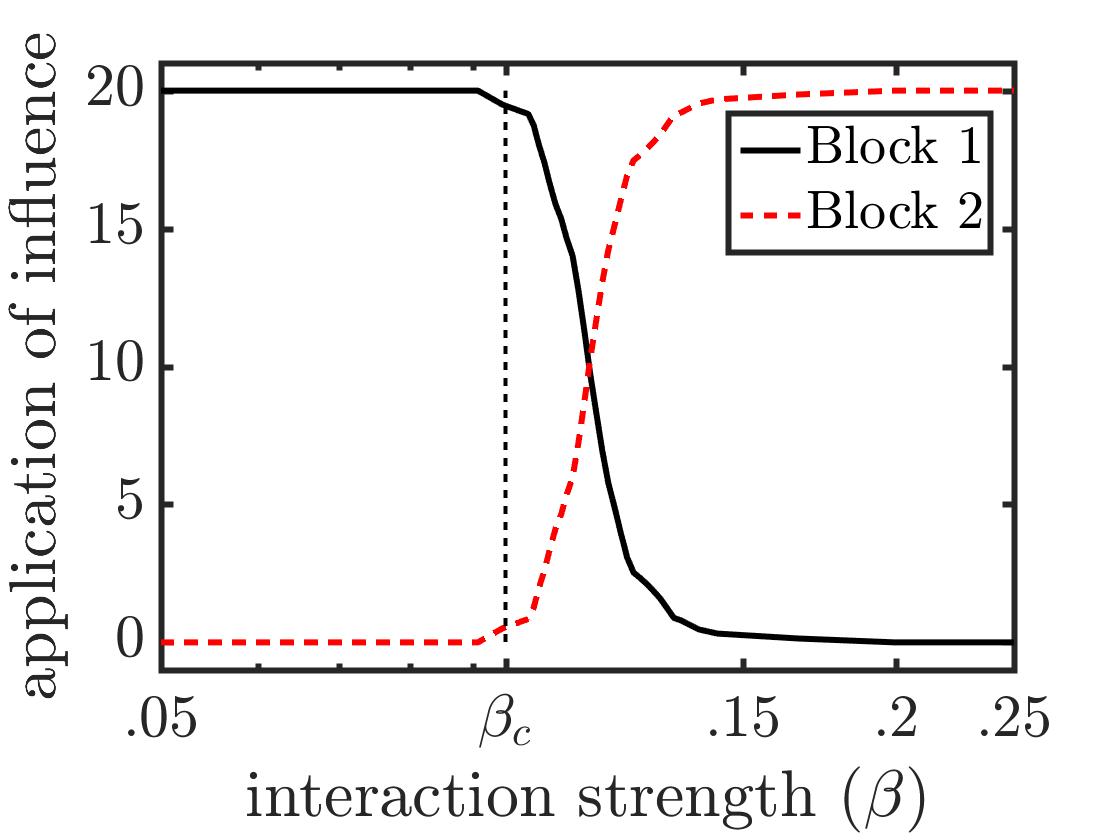}
\end{subfigure}%
\begin{subfigure}{0.4\textwidth}
        \includegraphics[width=\linewidth]{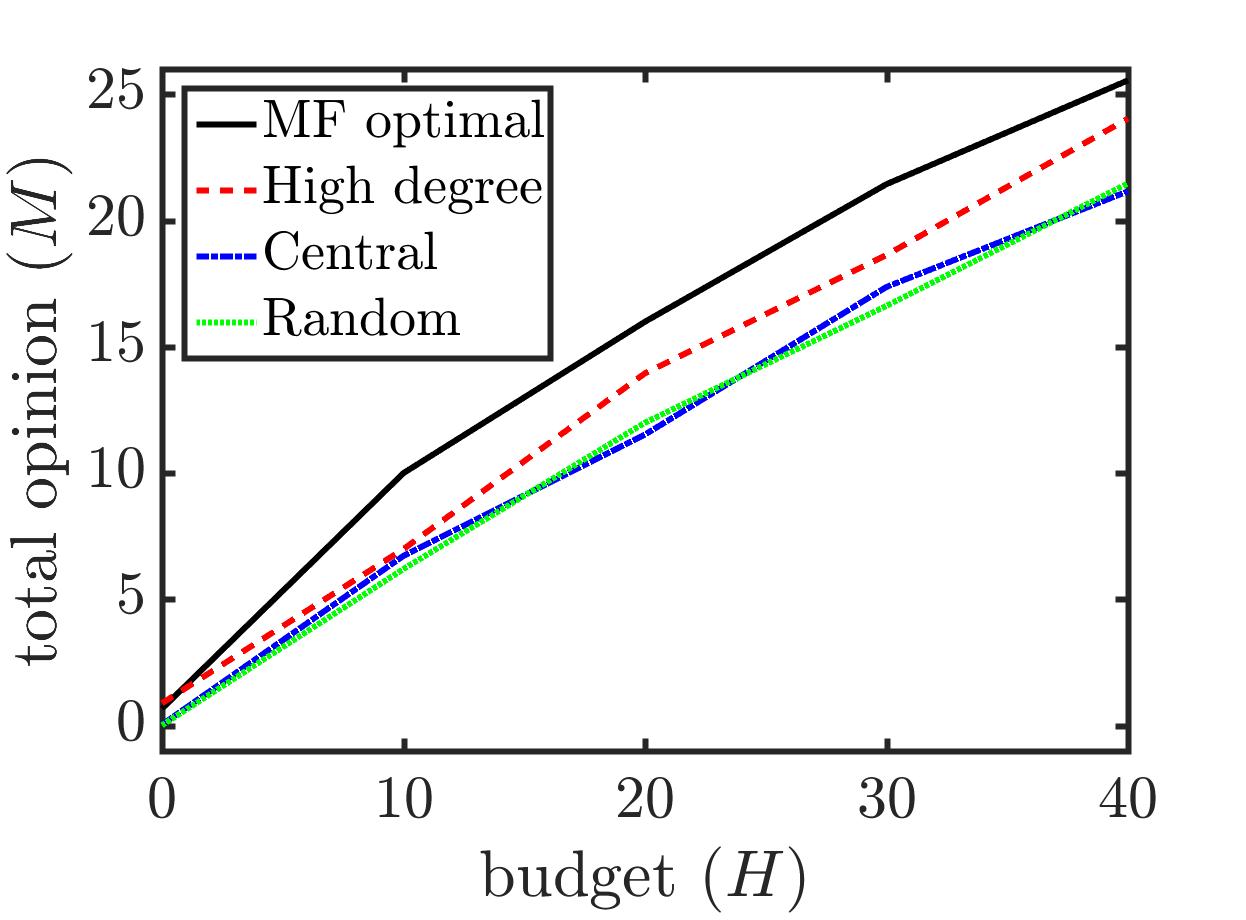}
\end{subfigure}
\caption{Left: A stochastic block network consisting of a highly-connected community (Block 1) and a sparsely-connected community (Block2). Center: The solution to MF-IIM shifts from focusing on Block 1 to Block 2 as $\beta$ increases. Right: Even at $\beta_c$, the MF solution outperforms common node-selection heuristics.}
\end{figure}
The stochastic block network is sufficiently large that exact calculation of the optimal external fields is infeasible. Thus, we resort to comparing the MF solutions with three node-selection heuristics: one that distributes the budget in amounts proportional to nodes' degrees, one that distributes the budget proportional to nodes' centralities (the inverse of a node's average shortest path length to all other nodes), and one that distributes the budget randomly. The magnetizations are approximated via Monte Carlo simulations of the Glauber dynamics, and we consider the system at $\beta=\beta_c$ to represent the worst-case scenario for the MF optimal external fields. The right plot in Figure 2 shows that, even at $\beta_c$, the solutions to MF-IIM outperform common node-selection heuristics. 

We consider a real-world collaboration network (Figure 3) composed of 904 individuals, where each edge is unweighted and represents the co-authorship of a paper on the arXiv \cite{SNAP-01}. We note that co-authorship networks are known to capture many of the key structural features of social networks \cite{Newman-01}. For $\bm{b}=0$ and $H=40$, the center plot in Figure 3 illustrates the sharp shift in the solution to MF-IIM at $\beta_c= 0.05$ from high- to low-degree nodes. Furthermore, the right plot in Figure 3 compares the performance of the MF optimal external field with the node-selection heuristics described above, where we again consider the system at $\beta_c$ as a worst-case scenario, demonstrating that Algorithm 1 is scalable and performs well on real-world networks.
\begin{figure}[t]
\centering
 \begin{subfigure}{0.2\textwidth}
        \includegraphics[width=\linewidth]{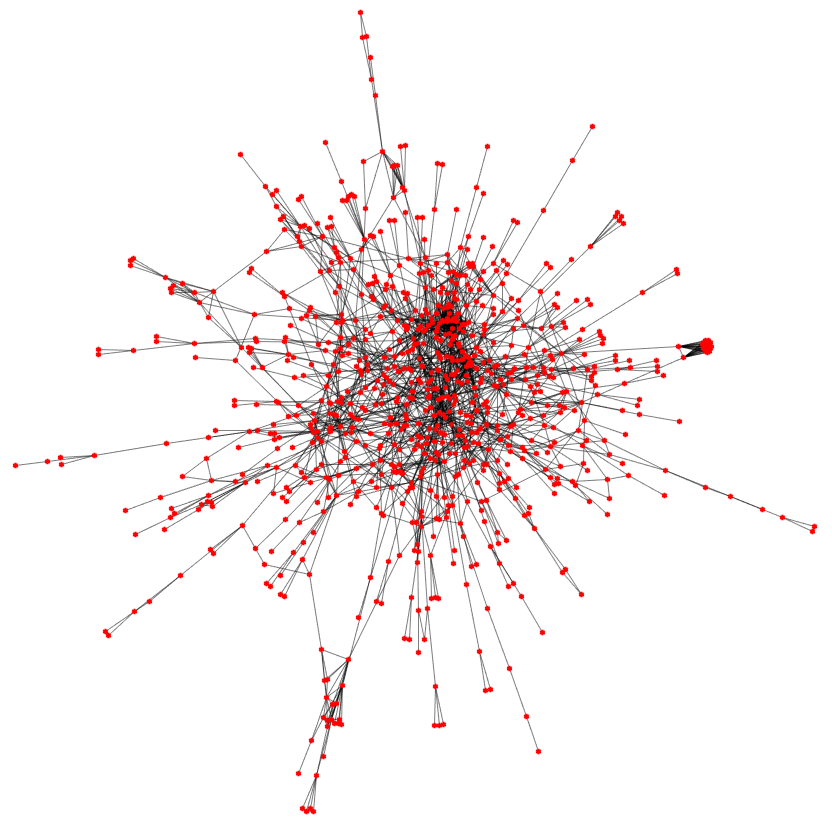}
\end{subfigure}%
\begin{subfigure}{0.4\textwidth}
        \includegraphics[width=\linewidth]{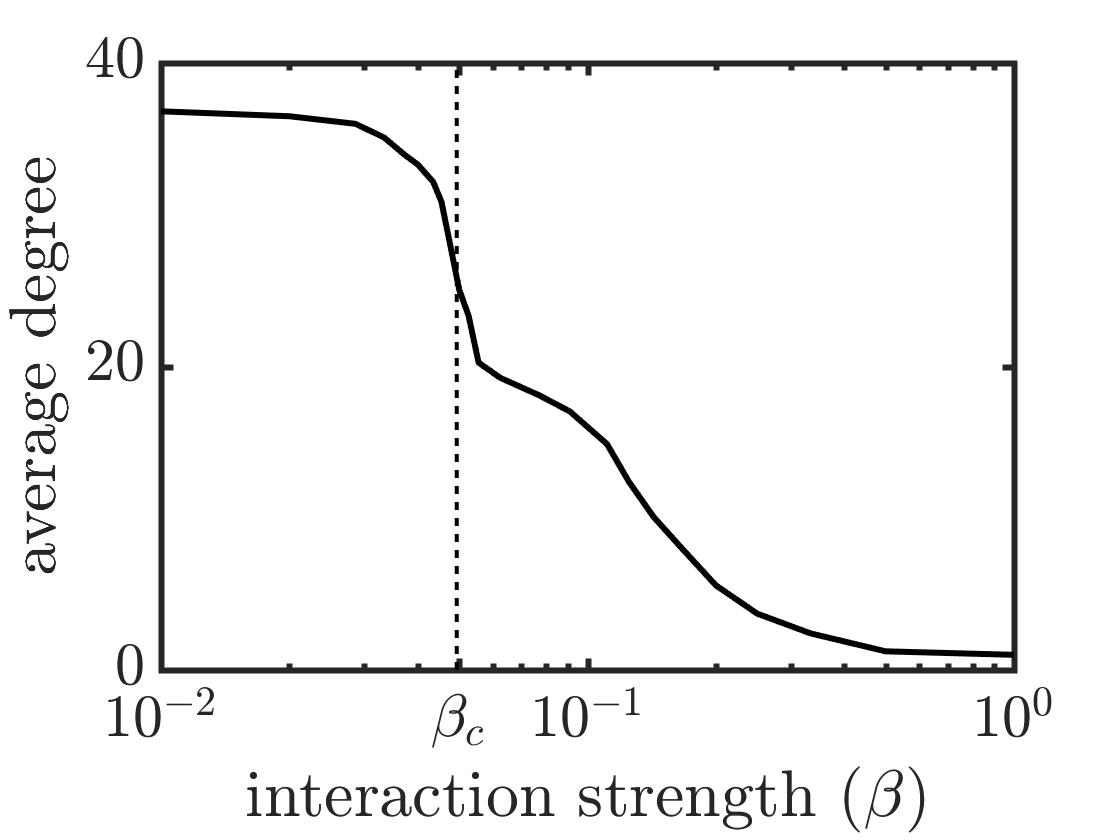}
\end{subfigure}%
\begin{subfigure}{0.4\textwidth}
        \includegraphics[width=\linewidth]{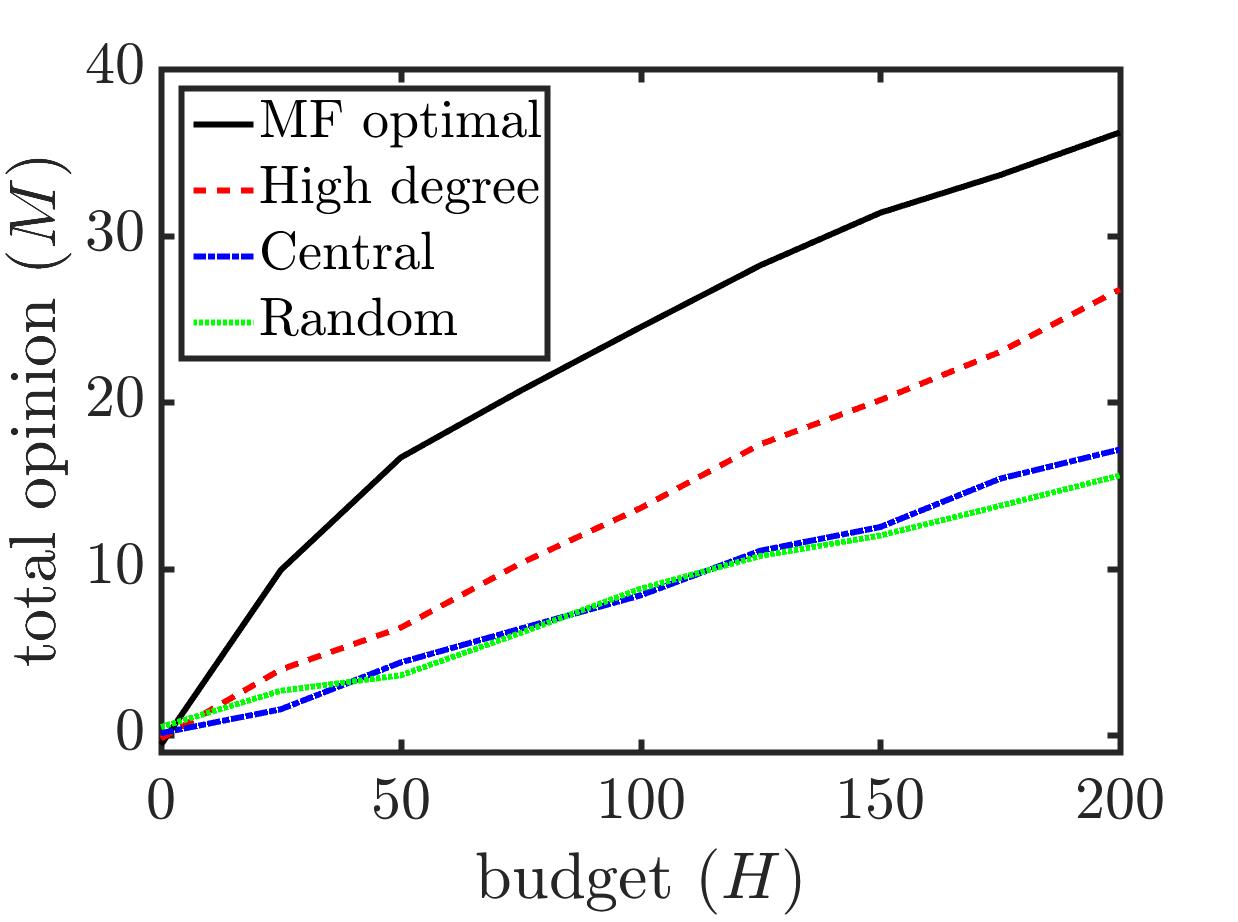}
\end{subfigure}
\caption{Left: A collaboration network of 904 physicists where each edge represents the co-authorship of a paper on the arXiv. Center: The solution to MF-IIM shifts from high- to low-degree nodes as $\beta$ increases. Right: The MF solution out-performs common node-selection heuristics, even at $\beta_c$.}
\end{figure}

\section{Conclusions}

We study influence maximization, one of the fundamental problems in network science, in the context of the Ising model, wherein repeated interactions between individuals give rise to complex macroscopic patterns. The resulting problem, which we call Ising influence maximization, has a natural physical interpretation as maximizing the magnetization of an Ising system given a budget of external magnetic field. Under the mean-field approximation, we develop a number of sufficient conditions for when the problem is concave, and we provide a gradient ascent algorithm that uses the susceptibility to efficiently calculate locally-optimal external fields. Furthermore, we demonstrate that the MF optimal external fields shift from focusing on high-degree individuals at low interaction strengths to focusing on low-degree individuals at high interaction strengths, a phenomenon not observed in viral models. We apply our algorithm on random and real-world networks, numerically demonstrating shifts in the solution structure and showing that our algorithm out-performs common node-selection heuristics.

It would be interesting to study the exact Ising model on an undirected network, in which case the spin statistics are governed by the Boltzmann distribution. Using this elegant steady-state description, one might be able to derive analytic results for the exact IIM problem. Our work establishes a fruitful connection between influence maximization and statistical physics, paving the way for exciting cross-disciplinary research. For example, one could apply advanced mean-field techniques, such as those in \cite{Yedidia-01}, to generate efficient algorithms of increasing accuracy. Furthermore, because our model is equivalent to a Boltzmann machine, one could propose a framework for data-based influence maximization based on well-known Boltzmann machine learning techniques.

\textbf{Acknowledgements.} We thank Michael Kearns and Eric Horsley for enlightening discussions, and we acknowledge support from the U.S. National Science Foundation, the Air Force Office of Scientific Research, and the Department of Transportation.
 
\bibliographystyle{unsrt}
\bibliography{IsingInfluenceBibNIPS}

\begin{thebibliography}{10}

\bibitem{Domingos-01}
P.~Domingos and M.~Richardson.
\newblock Mining the network value of customers.
\newblock {\em KDD}, pages 57--66, 2001.

\bibitem{Richardson-01}
M.~Richardson and P.~Domingos.
\newblock Mining knowledge-sharing sites for viral marketing.
\newblock {\em KDD'02. ACM}, pages 61--70, 2002.

\bibitem{Kempe-01}
D.~Kempe, J.~M. Kleinberg, and \'{E}. Tardos.
\newblock Maximizing the spread of influence through a social network.
\newblock {\em KDD'03. ACM}, pages 137--146, 2003.

\bibitem{Mossel-01}
E.~Mossel and S.~Roch.
\newblock On the submodularity of influence in social networks.
\newblock In {\em STOC'07}, pages 128--134. ACM, 2007.

\bibitem{Goyal-01}
S.~Goyal, H.~Heidari, and M.~Kearns.
\newblock Competitive contagion in networks.
\newblock {\em GEB}, 2014.

\bibitem{Gomez-Rodriguez-01}
M.~Gomez Rodriguez and B.~Sch{\"o}lkopf.
\newblock Influence maximization in continuous time diffusion networks.
\newblock In {\em ICML}, 2012.

\bibitem{Galam-03}
S.~Galam and S.~Moscovici.
\newblock Towards a theory of collective phenomena: consensus and attitude
  changes in groups.
\newblock {\em European Journal of Social Psychology}, 21(1):49--74, 1991.

\bibitem{Isenberg-01}
D.~J. Isenberg.
\newblock Group polarization: A critical review and meta-analysis.
\newblock {\em Journal of personality and social psychology}, 50(6):1141, 1986.

\bibitem{Mas-01}
M.~M{\"a}s, A.~Flache, and D.~Helbing.
\newblock Individualization as driving force of clustering phenomena in humans.
\newblock {\em PLoS Comput Biol}, 6(10), 2010.

\bibitem{Moussaid-01}
M.~Moussa{\"\i}d, J.~E. K{\"a}mmer, P.~P. Analytis, and H.~Neth.
\newblock Social influence and the collective dynamics of opinion formation.
\newblock {\em PLoS One}, 8(11), 2013.

\bibitem{De-01}
A.~De, I.~Valera, N.~Ganguly, S.~Bhattacharya, et~al.
\newblock Learning opinion dynamics in social networks.
\newblock {\em arXiv preprint arXiv:1506.05474}, 2015.

\bibitem{Montanari-01}
A.~Montanari and A.~Saberi.
\newblock The spread of innovations in social networks.
\newblock {\em PNAS}, 107(47), 2010.

\bibitem{Castellano-01}
C.~Castellano, S.~Fortunato, and V.~Loreto.
\newblock Statistical physics of social dynamics.
\newblock {\em Rev. Mod. Phys.}, 81:591--646, 2009.

\bibitem{Blume-01}
L.~Blume.
\newblock The statistical mechanics of strategic interaction.
\newblock {\em GEB}, 5:387--424, 1993.

\bibitem{McKelvey-01}
R.~McKelvey and T.~Palfrey.
\newblock Quantal response equilibria for normal form games.
\newblock {\em GEB}, 7:6--38, 1995.

\bibitem{Galam-02}
S.~Galam.
\newblock Sociophysics: a review of galam models.
\newblock {\em Int. J. Mod. Phys. C}, 19(3):409--440, 2008.

\bibitem{Sznajd-01}
K.~Sznajd-Weron and J.~Sznajd.
\newblock Opinion evolution in closed community.
\newblock {\em International Journal of Modern Physics C}, 11(06), 2000.

\bibitem{Kindermann-01}
R.~Kindermann and J.~Snell.
\newblock {\em Markov random fields and their applications}.
\newblock AMS, Providence, RI, 1980.

\bibitem{Tanaka-01}
T.~Tanaka.
\newblock Mean-field theory of boltzmann machine learning.
\newblock {\em PRE}, pages 2302--2310, 1998.

\bibitem{Nishimori-01}
H.~Nishimori and K.~M. Wong.
\newblock Statistical mechanics of image restoration and error-correcting
  codes.
\newblock {\em PRE}, 60(1):132, 1999.

\bibitem{Yedidia-01}
J.~Yedidia.
\newblock An idiosyncratic journey beyond mean field theory.
\newblock {\em Advanced mean field methods: Theory and practice}, pages 21--36,
  2001.

\bibitem{Jordan-01}
M.~I. Jordan, Z.~Ghahraman, T.~S. Jaakkola, and L.~K. Saul.
\newblock An introduction to variational methods for graphical models.
\newblock {\em Machine learning}, 37(2):183--233, 1999.

\bibitem{Opper-01}
M.~Opper and D.~Saad.
\newblock {\em Advanced mean field methods: Theory and practice}.
\newblock MIT press, 2001.

\bibitem{Saul-01}
L.~K. Saul, T.~Jaakkola, and M.~I. Jordan.
\newblock Mean field theory for sigmoid belief networks.
\newblock {\em Journal of artificial intelligence research}, 4(1):61--76, 1996.

\bibitem{Teboulle-01}
M.~Teboulle.
\newblock First order algorithms for convex minimization.
\newblock IPAM, 2010.
\newblock Tutorials.

\bibitem{SNAP-01}
J.~Leskovec and A.~Krevl.
\newblock {SNAP Datasets}: {Stanford} large network dataset collection, June
  2014.

\bibitem{Newman-01}
M.~Newman.
\newblock The structure of scientific collaboration networks.
\newblock {\em PNAS}, 98, 2001.

\end{thebibliography}

\end{document}


\maketitle

\section{Appendix A: Preliminaries}

We establish a number of preliminary results that aid the proofs of the theorems in the main text.

\subsection{Perron-Frobenius}

Many of the results in the paper rely on the Perron-Frobenius theorem, which we state here.

\textbf{Theorem 7. (Perron-Frobenius)}
\textit{Let $J$ be an irreducible non-negative matrix with spectral radius $\rho(J)=r$. Then the following statements hold:
\begin{enumerate}
\item $J$ has a real, positive, and simple eigenvalue equal to $r$.
\item The corresponding eigenvector of $J$ has all-positive components.
\item If $0\le J \le A$, for some matrix $A$, then $r_J\le r_A$.
\end{enumerate}
}

It is known that the adjacency matrix of a strongly-connected graph is irreducible, and hence, all of the results of the Perron-Frobenius theorem carry over.

\subsection{The existence of a unique and stable steady-state for $\beta<\beta_c$}

We first show that if our network is strongly-connected, then for $\beta < 1/\rho(J)$, the system exhibits a unique steady-state that is stable under $\bm{f}$. This result will aid in the proof of Theorem 3 and similar arguments are used in the proof of Theorem 4.  The proof in this section is based on Banach's fixed point theorem, but instead of directly showing that $\bm{f}$ is a contraction mapping on $X=[-1,1]^n$, we use the spectral properties of $\bm{f}'$. The following two lemmas relate the contraction mapping property to the spectral radius of $\bm{f}'$. We note that throughout the proofs, we use the variable $\bm{x}$ instead of $\bm{m}$ to indicate a point in $X$ that is not necessarily a steady-state of $\bm{f}$.

\textbf{Lemma 8.}
\textit{Let $X$ be a convex subset of Euclidean space and let the function $\bm{f}:X\rightarrow X$ have continuous partial derivatives on $X$. If the Jacobian satisfies
\begin{equation}
|\bm{f}'(\bm{x})|<1,
\end{equation}
for all $\bm{x}\in X$ and some matrix norm $|\cdot|$, then $\bm{f}$ satisfies the contraction mapping property on $X$.}

\textbf{Lemma 9.}
\textit{Given a square matrix $A$ and $\epsilon > 0$, there exists a matrix norm $|\cdot|$ such that
\begin{equation}
|A|\le \rho(A) + \epsilon.
\end{equation}}

We are now prepared to show that if $J$ is strongly-connected, then for $\beta<1/\rho(J)$, $\bm{f}$ is a contraction mapping on $X$, and hence $\bm{f}$ admits a unique and stable fixed point on $X$.

\textbf{Lemma 10.}
\textit{Let $(N,J,\bm{h},\beta)$ describe a system with a strongly-connected graph. For $\beta<\beta_c$, there exists a unique and stable steady-state that can be found by iteratively applying $\bm{f}$ to any point $\bm{x}\in X$.}

\textit{Proof.}
Consider the Jacobian,
\begin{equation}
\bm{f}'(\bm{m})_{ij} = \frac{\partial}{\partial m_j} \tanh\left[\beta(J\bm{m}+\bm{h})\right]_i = \beta \text{sech}^2 \left[\beta(J\bm{m}+\bm{h})\right]_i J_{ij}.
\end{equation}
Since $|\text{sech}(\cdot)|\le 1$, $|\bm{f}'(\bm{m})_{ij}|\le \beta J_{ij}$ for all $i,j\in N$. For $\beta<1/\rho(J)$ and for all $\bm{m}\in X$, we have
\begin{equation}
\rho(\bm{f}'(\bm{m})) \le \rho(\beta J) = \beta \rho(J) < 1,
\end{equation}
where the first inequality follows from the Perron-Frobenius theorem and the equality follows from the linearity of $\rho(\cdot)$.

Because the above inequality is strict, there exists an $\epsilon>0$ such that $\rho(\bm{f}'(\bm{x}))+\epsilon < 1$ for all $\bm{x}\in X$. By Lemma 9 we can choose a matrix norm $|\cdot|$ such that
\begin{equation}
|\bm{f}'(\bm{m})|\le \rho(\bm{f}'(\bm{m})) + \epsilon < 1.
\end{equation}
Since $X$ is a convex subset of Euclidean space, Lemma 8 implies that $\bm{f}$ satisfies the contraction mapping property on $X$. Since $X$ is a closed and bounded it is also a compact metric space and we can apply Banach's theorem on compact metric spaces to attain the desired result. \hfill $\square$

\subsection{The smoothness of stable non-negative steady-states for increasing $h$}

We show that any stable non-negative steady-state $\bm{m}$ gives rise to a unique and stable branch of steady-states that is smooth and non-decreasing as $\bm{h}$ increases. We note that this result represents half of the progress toward proving Theorem 5.

\textbf{Lemma 11.}
\textit{Let $(N,J,\bm{h},\beta)$ describe a system with a strongly-connected graph for which there exists a stable steady-state $\bm{m}$. If $\bm{m}\ge 0$, then as $\bm{h}$ increases, $\bm{m}$ gives rise to a unique and smooth branch of stable non-negative steady-states.}

\textit{Proof.}
We first show that any stable steady-state is locally non-decreasing in $\bm{h}$. Consider the susceptibility from Sec. 2:
\begin{equation}
\chi^{MF} = \beta(I-\beta D(\bm{m})J)^{-1}D(\bm{m}) =  \beta(I-\bm{f}'(\bm{m}))^{-1}D(\bm{m}).
\end{equation}
Since $\rho\left(\bm{f}'(\bm{m})\right) < 1$, Theorem 4.3 of \cite{Fielder-01} guarantees that the matrix $(I-\bm{f}'(\bm{m}))^{-1}$ is non-negative. Furthermore, since $D(\bm{m})$ is non-negative, we find that $\chi^{MF}$ is non-negative, and hence $\frac{\partial m_i}{\partial h_j}\ge 0$ for all $i,j\in N$.

We now argue that $\bm{m}$ remains stable as $\bm{h}$ increases which, by the implicit function theorem, guarantees that $\bm{m}$ branches uniquely and smoothly. It is sufficient to show that $\rho(\bm{f}'(\bm{m})) = \rho(\beta D(\bm{m})J)$ is non-increasing in $\bm{h}$. Since $\bm{m}$ is non-decreasing in $\bm{h}$, we find that $\bm{f}'(\bm{m})$ is entry-wise non-increasing in $\bm{h}$, and hence Perron-Frobenius guarantees that $\rho(\bm{f}'(\bm{m}))$ is non-increasing. \hfill $\square$

\section{Appendix B: Proofs}

We now present proofs of the theorems in the main text, noting that we do not present the results in the order that they appear in the text since some earlier results depend on later ones.

\subsection{Theorem 4}

We split Theorem 4 into three separate results. We first show that any stable non-negative steady-state is the unique stable non-negative steady state and that it attains the maximum total opinion among all steady-states. Secondly, we note that Lemma 11 guarantees that any stable non-negative steady-state is smooth and remains stable and non-negative for increasing $\bm{h}$. Finally, we show that any stable non-negative steady-state is concave in $\bm{h}$. Together, these results prove Theorem 4.

\subsubsection{Uniqueness of stable non-negative steady-states}

We show that any stable non-negative steady-state is the unique stable non-negative steady-state and achieves the largest total opinion among steady-states. First consider the following lemma.

\textbf{Lemma 12.}
\textit{Let $(N,J,\bm{h},\beta)$ describe an arbitrary system and consider any point $\bm{x}\in X$. Then
\begin{equation}
\bm{f}^{\ell}(\bm{1}) \ge \bm{f}^{\ell}(\bm{x}),
\end{equation}
for any positive integer $\ell$, where $\bm{f}^{\ell}(\cdot)$ denotes the $\ell^{th}$ iterative application of $\bm{f}$ and $\bm{1}$ is the vector of ones of length $n$.}

\textit{Proof.}
We proceed by induction. The base case is trivially satisfied. For the inductive step, assume $f^{\ell}(\bm{1})_i\ge f^{\ell}(\bm{x})_i$ for some $\ell$ and all $i\in N$. Since $J\ge 0$, $\beta\ge 0$ and $\tanh(\cdot)$ is increasing, we have
\begin{equation}
f^{\ell+1}(\bm{1})_i = \tanh\left[\beta\left(Jf^{\ell}(\bm{1})+\bm{h}\right)\right]_i \ge \tanh\left[\beta\left(Jf^{\ell}(\bm{x})+\bm{h}\right)\right]_i = f^{\ell+1}(\bm{x})_i,
\end{equation}
for all $\bm{x}\in X$ and all $i\in N$. \hfill $\square$

We now establish the uniqueness of stable non-negative steady-states.

\textbf{Lemma 13.}
\textit{Let $(N,J,\bm{h},\beta)$ describe a system with a strongly-connected network for which there exists a stable non-negative steady-state $\bm{m}$. Then $\bm{m}$ is the unique stable steady-state and can be found by iteratively applying $\bm{f}$ to $\bm{1}$.}

\textit{Proof.}
Assume there exists a stable non-negative steady-state $\bm{m}$. By Lemma 12,
\begin{equation}
\bm{f}^{\ell}(\bm{1}) \ge \bm{f}^{\ell}(\bm{m}) = \bm{m},
\end{equation}
for any $\ell$. This indicates that the sequence $\{ \bm{f}^{\ell}(\bm{1})\}$ is contained in the closed region $X_{\bm{m}}=\{\bm{x}\in X : \bm{x}\ge \bm{m}\}$. By an argument similar to that in the proof of Lemma 11, we have $\rho(\bm{f}'(\bm{x}))\le \rho(\bm{f}'(\bm{m})) < 1$ for all $\bm{x}\in X_{\bm{m}}$. By an argument similar to that in the proof of Lemma 10, this indicates that $\bm{f}$ is a contraction mapping on $X_{\bm{m}}$.

Following the proof in \cite{Palais-01}, we show that the sequence $\{ \bm{f}^{\ell}(\bm{1})\}$ is Cauchy. Since $|\bm{f}(\bm{x}')-\bm{f}(\bm{x})|< |\bm{x}' - \bm{x}|$ for all $\bm{x},\bm{x}'\in X_{\bm{m}}$, there exists a number $q\in(0,1)$ such that $|\bm{f}(\bm{x}') - \bm{f}(\bm{x})|\le q |\bm{x}' -\bm{x}|$. By the triangle inequality,
\begin{equation}
|\bm{x}' - \bm{x}| \le |\bm{x}' - \bm{f}(\bm{x}')| + q|\bm{x}' - \bm{x}| + |\bm{f}(\bm{x}) - \bm{x}|, 
\end{equation}
which yields
\begin{equation}
|\bm{x}' - \bm{x}|\le \frac{|\bm{f}(\bm{x}') - \bm{x}'| + |\bm{f}(\bm{x}) - \bm{x}|}{1-q}.
\end{equation}
Replacing $\bm{x}$ and $\bm{x}'$ with $\bm{f}^{\ell}(\bm{1})$ and $\bm{f}^k(\bm{1})$, respectively, we find
\begin{align}
|\bm{f}^k(\bm{1}) - \bm{f}^{\ell}(\bm{1})| &\le \frac{|\bm{f}^{k+1}(\bm{1}) - \bm{f}^k(\bm{1})| + |\bm{f}^{\ell+1}(\bm{1}) - \bm{f}^{\ell}(\bm{1})|}{1-q} \\
&\le \frac{q^k + q^{\ell}}{1-q}|\bm{f}(\bm{1}) - \bm{1}|. \nonumber
\end{align}
Since $q<1$, the last expression goes to zero as $\ell,k\rightarrow\infty$, proving that $\{ \bm{f}^{\ell}(\bm{1})\}$ is Cauchy and hence converges to a limit $\bm{m}^*\in X_{\bm{m}}$. Furthermore, the limit $\bm{m}^*$ is a fixed point of $\bm{f}$, and hence a steady-state of the system, since
\begin{equation}
\bm{m}^* = \lim_{\ell\rightarrow\infty} \bm{f}^{\ell}(\bm{1}) = \lim_{\ell\rightarrow\infty} \bm{f}(\bm{f}^{\ell-1}(\bm{1})) = \bm{f}\left(\lim_{\ell\rightarrow\infty} \bm{f}^{\ell-1}(\bm{1})\right) = \bm{f}(\bm{m}^*).
\end{equation}

Suppose for contradiction that $\bm{m}^*\neq \bm{m}$, and consider the line $(1-t)\bm{m} + t\bm{m}^*$ between $\bm{m}$ and $\bm{m}^*$ for $t\in[0,1]$. All points along this line lie in $X_{\bm{m}}$ and hence $\bm{f}$ is contractive along the line. We have,
\begin{align}
\left| \bm{f}(\bm{m}^*)-\bm{f}(\bm{m})\right| &= \left| \int_{\bm{m}}^{\bm{m}^*} \bm{f}'(\bm{x})\cdot d\bm{x}\right| \\
&\le  \int_0^1\left| \bm{f}'\left((1-t)\bm{m} + t\bm{m}^*\right)\right| \left| \bm{m}^*-\bm{m} \right| dt, \nonumber
\end{align}
where $|\bm{f}'(\cdot)|$ represents any matrix norm. Because $\bm{f}$ is contractive along the line, we can choose a matrix norm that is strictly less than 1. Thus,
\begin{equation}
\left| \bm{f}(\bm{m}^*)-\bm{f}(\bm{m})\right|< \int_0^1 \left| \bm{m}^*-\bm{m} \right| dt = \left| \bm{m}^*-\bm{m}\right|,
\end{equation}
which is a contradiction. Thus $\bm{m}^*=\bm{m}$ and the stable non-negative steady-state is unique. Furthermore, this shows that $\{ \bm{f}^{\ell}(\bm{1})\}$ converges to $\bm{m}$. \hfill $\square$

As a corollary, we find that any stable non-negative steady-state attains the maximum total opinion among all steady-states.

\textbf{Corollary 14.}
\textit{Let $(N,J,\bm{h},\beta)$ describe a system for which there exists a stable non-negative steady-state $\bm{m}$, and let $\bm{m}'$ be another steady-state. Then $\bm{m}\ge \bm{m}'$.}

\textit{Proof.}
By Lemmas 12 and 13 we have
\begin{equation}
\bm{m} = \lim_{\ell\rightarrow\infty} \bm{f}^{\ell}(\bm{1}) \ge \lim_{\ell\rightarrow\infty} \bm{f}^{\ell}(\bm{m}')=\bm{m}',
\end{equation}
for any steady-state $\bm{m}'$. \hfill $\square$

\subsubsection{The concavity of stable non-negative steady-states in $h$}

We show for $J$ strongly-connected that any stable non-negative steady-state is concave in $\bm{h}$.

\textbf{Lemma 15.}
\textit{Let $(N,J,\bm{h},\beta)$ describe a system with a strongly-connected graph for which there exists a stable non-negative steady-state $\bm{m}$. Then $\bm{m}$ is concave in $\bm{h}$.}

\textit{Proof.}
We want to show that the Hessian of $m_i$ with respect to $\bm{h}$ is negative semidefinite for all $i\in N$. The Hessian of $m_i$ with respect to $\bm{h}$ is given by
\begin{equation}
C_{jk}^{(i)}\equiv \frac{\partial^2 m_i}{\partial h_j \partial h_k} = \frac{\partial}{\partial h_k}\chi_{ij}^{MF}.
\end{equation}
After taking partials and rearranging we are left with
\begin{equation}
C_{jk}^{(i)} = -2\sum_{\ell\in N} {\chi_{j\ell}^{MF}}^T\left(\frac{m_{\ell}}{(1-m_{\ell}^2)^2}\chi_{i\ell}^{MF}\right)\chi_{\ell k}^{MF} = -\sum_{\ell\in N} Z_{j\ell}^{(i)T} Z_{\ell k}^{(i)},
\end{equation}
where $Z_{jk}^{(i)}= \chi_{jk}^{MF}\sqrt{\frac{2m_j}{(1-m_j^2)^2}\chi_{ij}^{MF}}$. Since $m_j,\chi_{ij}^{MF}\ge 0$ for all $i,j\in N$, $Z^{(i)}$ is a real matrix. Thus $C^{(i)}$ is negative semidefinite for all $i\in N$. \hfill $\square$

\subsection{Theorem 5}

We show that any stable non-negative steady-state $\bm{m}$ gives rise to a unique and stable branch of steady-states that is smooth and non-decreasing as $\bm{h}$ increases. We also show that if $\bm{m}>0$, then $\bm{m}$ gives rise to a unique and stable branch of steady-states that is smooth and non-decreasing as $\beta$ increases. We note that the first result is given by Lemma 11. To prove the second result, we first show that any stable positive steady-state is locally non-decreasing in $\beta$.

\textbf{Lemma 16.}
\textit{Let $(N,J,\bm{h},\beta)$ describe any system for which there exists a stable positive steady-state $\bm{m}$. Then $\bm{m}$ is locally non-decreasing in $\beta$.}

\textit{Proof.}
We want to show $\frac{d m_i}{d \beta}$ is non-negative for all $i\in N$. By assumption, $\rho\left(\bm{f}'(\bm{m})\right) < 1$, allowing us to apply the implicit function theorem, giving
\begin{align}
\frac{d m_i}{d \beta} &= \sum_{j\in N} \left(\delta_{ji} - \bm{f}'(\bm{m})_{ji}\right)^{-1} \frac{\partial f_j}{\partial \beta} \\
&= \sum_{j\in N} \left(\delta_{ji} - \bm{f}'(\bm{m})_{ji}\right)^{-1} \text{sech}^2\left[\beta(J\bm{m}+\bm{h})\right]_j \left(J\bm{m}+\bm{h}\right)_j. \nonumber
\end{align}
In vector form,
\begin{equation}
\label{DmDb}
\frac{d \bm{m}}{d\beta} = (I - \bm{f}'(\bm{m}))^{-1}D(\bm{m})(J\bm{m}+\bm{h}).
\end{equation}
Theorem 4.3 of \cite{Fielder-01} guarantees that the matrix $(I-\bm{f}'(\bm{m}))^{-1}$ is non-negative and $D(\bm{m})$ is also non-negative. Because $\bm{m}=\tanh\left[\beta(J\bm{m}+\bm{h})\right]> 0$, we have $J\bm{m}+\bm{h}> 0$, and hence Eq. (\ref{DmDb}) is non-negative. \hfill $\square$

We now complete the proof of Theorem 5, showing that any stable positive steady-state gives rise to a unique and stable branch of steady-states as $\beta$ increases.

\textit{Proof (Theorem 5).}
For contradiction, assume that increasing $\beta$ causes $\bm{m}$ to lose stability. Because the network is strongly-connected, $\bm{f}'(\bm{m}) = \beta D(\bm{m})J$ is also strongly-connected. Thus, Perron-Frobenius guarantees that $\bm{f}'(\bm{m})$ has a simple largest eigenvalue equal to its spectral radius. When $\bm{m}$ loses stability, this simple eigenvalue crosses one from below. By the Crandall-Rabinowitz theorem \cite{Crandall-01} and the principle of exchange of stability, the crossing of the simple eigenvalue gives rise to two new sable steady-states. However, Lemma 16 guarantees that $\bm{m}$ remains positive as we increase $\beta$, which necessitates that both of the new stable steady-states are also initially positive, contradicting Theorem 4. Thus, $\bm{m}$ cannot lose stability as $\beta$ increases, and hence $\bm{m}$ gives rise to a unique and smooth branch of stable and positive steady-states. \hfill $\square$

\subsection{Theorem 3}

We show that every system exhibits a steady-state and that the well-known pitchfork bifurcation structure for steady-states of the ferromagnetic MF Ising model on a lattice extends exactly to general (weighted, directed) strongly-connected graphs. In particular, for any strongly-connected graph $J$, there is a critical interaction strength $\beta_c = 1/\rho(J)$ below which there exists a unique and stable steady-state. For $\bm{h}=0$, as $\beta$ crosses $\beta_c$ from below, two new stable steady-states appear, one with all-positive components and one with all-negative components.

\textit{Proof (Theorem 3).}
We first note that the existence of a steady-state is guaranteed for any system by applying Brouwer's fixed point theorem to $\bm{f}$. Furthermore, Lemma 10 establishes that for $\beta<1/\rho(J)$, there is a unique and stable steady-state.

In the case $\bm{h}=0$, any system has a steady-state at $\bm{m}^*=0$, which we refer to as the \textit{trivial steady-state}. Lemma 10 guarantees that $\bm{m}^*$ is stable and unique for $\beta<\beta_c$. The implicit function theorem guarantees that we can continue to write $\bm{m}^*$ uniquely as a function of $\beta$ so long as $\rho(\bm{f}'(\bm{m}^*))=\beta\rho(J)<1$. If our network is strongly-connected, then the Perron-Frobenius theorem guarantees that as we increase $\beta$, an eigenvalue of $\bm{f}'$ will first cross 1 when $\beta=1/\rho(J)$. Furthermore, the largest eigenvalue is simple, which, by the Crandall-Rabinowitz theorem \cite{Crandall-01}, guarantees the appearance of two new steady-states. Furthermore, the new solutions locally lie in the subspace spanned by the eigenvector corresponding to the largest eigenvalue of $\bm{f}'$, which by the Perron-Frobenius theorem has all positive components. Thus, at $\beta=\beta_c$, a branch of steady-states appears, giving rise to an all-positive steady-state and an all-negative steady-state. By the principle of exchange of stability, the new steady-states adopt the stability of the trivial steady-state, while the trivial steady-state becomes unstable. As we continue to increase $\beta$, Theorem 5 guarantees that the positive (negative) steady-state remains positive (negative) and stable. \hfill $\square$

\subsection{Theorem 6}

We conclude by showing for $J$ strongly-connected that if $\bm{h} \ge 0$, then there exists a stable non-negative steady-state.

\textit{Proof (Theorem 6).}
We first consider $\bm{h}>0$. Lemma 10 guarantees that for any $\beta<\beta_c$ there exists a unique and stable steady-state $\bm{m}$ and that iterative application of $\bm{f}$ to any $\bm{x}\in X$ converges to $\bm{m}$. For induction, choose $\bm{x}=\bm{1}$ and assume $\bm{f}^{\ell}(\bm{1})>0$. Then
\begin{equation}
\bm{f}^{\ell+1}(\bm{1})=\tanh\left[\beta(J\bm{f}^{\ell}(\bm{1})+\bm{h})\right]>0.
\end{equation}
Thus, $\bm{m} = \lim_{\ell\rightarrow\infty} \bm{f}^{\ell}(\bm{1})>0$ at $\beta$. By Theorem 5, the unique branch $\bm{m}(\beta)$ remains stable and positive for all $\beta'>\beta$. To complete the proof, we note that Theorem 3 covers the case $\bm{h}=0$. \hfill $\square$
  
\bibliographystyle{abbrvnat}
\bibliography{IsingInfluenceBibNIPS}